\documentstyle[epsfig]{aipproc}

\begin{document}
\title{A Search for Gamma-Ray Burst Optical Emission with the
Automated Patrol Telescope}

\author{Bruce Grossan$^*$, Saul Perlmutter$^*$ and Michael Ashley$^{\dagger}$}
\address{$^*$University of California at Berkeley, Lawrence Berkeley
National Laboratory \\
$^{\dagger}$University of New South Wales Dept. of Physics and
Astronomy\\
}

To be published in ``Gamma-Ray Bursts, 4th Huntsville Symposium'',
1998, ed. C. Meegan, R. Preece, T. Koshut (New York:AIP)

\maketitle

\begin{abstract}
The Automated Patrol Telescope (APT) is a wide-field ($5^\circ \times 5^\circ$), modified 
Schmidt capable of covering large gamma-ray burst (GRB) localization 
regions to produce 
a high rate of GRB optical emission measurements.  Accounting 
for factors such as bad weather and incomplete overlap of our 
field and large GRB localization regions, we estimate our search 
will image the actual location of 20-41 BATSE GRB sources each 
year. %
{\bf  }Long exposures will be made for these images, repeated for
several nights, to detect delayed optical transients (OTs) with light 
curves similar to those already discovered.  The APT can also 
respond within about 20 sec. to GRB alerts from BATSE to 
search for prompt emission from GRBs.  We expect 
to image more than 2.4 GRBs  yr.$^{-1}$ during $\gamma$-ray emission. 
 More than 5.1 will be imaged yr.$^{-1}$ within $\sim$20 sec. 
of emission.  The APT's 50 cm aperture is much larger than other 
currently operating experiments used to search for prompt emission, 
and the APT is the only GRB dedicated telescope in the Southern 
Hemisphere.  Given the current rate of $\sim$25\% OTs per X/$\gamma$ 
localization, we expect to produce a sample of $\sim$10 OTs for 
detailed follow-up observations in 1-2 years of operation.
\end{abstract}

\section*{Introduction}
%
%
%
%
This meeting featured observations of the two optical transients (OTs) 
associated with gamma-ray bursts (GRBs) so far.  These important 
observations have produced some controversial results (see presentations 
by Caraveo, Lamb, and Fruchter), and they have not yet yielded 
either explanations of the physics of the bursts or an identification 
of the bursting object(s).  An obvious next step is to find and 
study a %
{\it sample} of OTs, not just two examples as we have now.  
Observations of $\sim$10 or more OTs are needed to address the 
controversies and give a statistical basis to host galaxy incidence 
and other properties. 

We have begun to use the wide-field ($5^\circ \times 5^\circ$)
Automated Patrol Telescope (APT) to image GRB 
positions for the identification of GRB-associated OTs.  The
telescope, located at Siding Spring Observatory in Australia, is fully
dedicated to this project.  The wide field is essential to search the
large GRB localizations from BATSE, the majority of all GRB
localizations.  In one mode of operation, we will study delayed
emission similar to the OTs already discovered. 
In a second mode of operation, we will search for prompt 
optical emission, actually during or shortly after gamma-ray emission.

\section*{Project Description}
Our telescope 
performs a regular observing program unless a GRB alert is received. 
 The Gamma-Ray Coordinates Network (GCN) sends out alerts
1-3 sec. after a BATSE GRB reaches threshold; these alerts 
reach our site via the Internet in 200-400 msec.  The
APT then interrupts its observations, slews to the 
GRB position, and begins integration.  %

{\bf 1) Prompt Emission Search: } Because the APT can make long 
slews in $<$ 20 sec., we can make optical observations during 
the tail end of $\gamma$-ray emission for many GRBs.  Figure 1 shows 
that more than 47\% of all bursts are longer than our 20 sec. 
slew time.  Shorter bursts would still be observed less than 
20 sec. after $\gamma$ emission.  

After the initial slew, a series 
of exposures of 10, 20, 40, ... sec. are taken to sample different 
time scales up to $\sim$2 hrs.  Our CCD can be read 
in 6-10 sec.  We expect to reach a sensitivity of V$>$17.7 
mag at 5 $\sigma$ in 10 sec. exposures.  

{\bf 2) Delayed Emission Search:}  
For up to 4 nights after the GRB (double the time-to-peak for OT970508), 
1 hr. of exposures are acquired every other hour the source 
position can be observed.  In this way, we will sample light 
curves at good time resolution and build up long total exposure 
times, but still do some scheduled observing.  The APT will reach a
sensitivity of V$>$20.9 at 5 $\sigma$ in 1 hr. of co-added
exposures.  Fig. 2 shows that OT970508 would be detectable by the APT at better 
than 5$\sigma$ in less than 1/2 hr. of exposure.  

We will search for OTs by subtracting images taken just after the 
alert, and those taken later at the same position.  Automated 
software will then examine the subtractions for transients, measure
their position and brightness, and eliminate
false candidates.   To facilitate follow-up, our positions (better
than $\sim$ $5^{\prime\prime}$) will be rapidly publicized using the GCN e-mail list.

\subsection*{Status}
During March - April of 1997, a short pilot search was undertaken to 
demonstrate and test our system.  While the APT was performing 
unrelated observations, and while no operator 
was present, the system responded to GRBs 970326b 
and 970329.  The system worked as planned, producing a series 
of images at the alert locations for 
both bursts; one of them is shown in Figure 3.  (The BACODINE 
original positions for these two ``test'' events were of very 
poor accuracy, and the GRB position was unlikely to be
on our single-pointing fields.)   During another run,
GRB 970616 was observed for $>$1 hr.  
At this time, the APT requires an operator to 
open and close the telescope each night. 
Also, our CCD covers only a $3^\circ \times  2^\circ$ field at 
a quantum efficiency of only 28\% in V.  To implement our full 
search, we plan to upgrade the camera with a 79\% quantum efficiency 
(at V) 2048 $\times$ 2048 detector to cover our $5^\circ$ square field.
We will also automate the telescope open/closing 
procedure and weather monitoring to enable unsupervised searching 
every clear night.

\subsection*{Event Rate}
Important improvements are expected in the GCN which will provide LOCBURST 
quality positions ($4^\circ$-$11^\circ$ 68\% error diameter, see Kippen et 
al. and Briggs et al., this volume) as soon as most of the photons 
in a burst are received.  These ``updated positions'' are expected to be in 
place by early 1998.  Our ``prompt emission search'' 
event rate was calculated by taking into account time lost due 
to weather (35\% for our site), and time lost due to the moon increasing 
our background by more than a mag per square arc sec ($\sim$5\% 
of the time not already counted), and the fraction 
(48\%) of GRBs longer than 20 sec. (our estimated average time 
before pointing).  Of the 27 updated positions per year for our
site, we calculate that our system will slew to about 18 bursts per
year, more than 8 longer than 20 sec.  Taking into account the
overlap of our field with the updated position errors, %
{\bf our prompt emission search will image } %
{\bf more than 2.4 GRBs yr.}%
{\bf $^{-1}$ }%
{\bf  during }%
{\bf $\gamma$-}%
{\bf ray emission,} %
{\bf  and more than 5.1 yr.$^{-1}$ }%
{\bf }%
{\bf  within $\sim$20 sec. of emission.}  (We use ``image 
a GRB'' here to mean that the actual position of the GRB is on 
the detector, resulting in either a measurement or an upper limit.)

Our ``delayed emission search'' event rate was calculated by taking 
into account weather and moon as above, 
but considering GRBs of all durations, 
and all of the sky accessible any time during the night ($\sim$64\%). 
From 52-104 BATSE LOCBURST positions yr.$^{-1 }$, %
{\bf  we estimate our delayed emission search will image 20-41 
LOCBURST sources yr.$^{-1 }$} by covering the larger error boxes 
with 2 $\times$ 2 mosaics.  %
We will also acquire 4.7 images from RXTE ASM alerts, and 3.5 images 
from SAX alerts each year.

\subsection*{Comparison To Other Efforts}
The LOTIS and ROTSE are currently operating, rapid-slewing, wide-field
11 cm aperture instruments  
custom built to search for prompt emission.  They have faster response
and a higher event rate, however, the APT's 50 cm
aperture yields much greater sensitivity.  LONEOS  (see Wagner 
\& Shrader, this work) has a comparable aperture to the APT, 
but has a limited commitment to GRB observations, and lacks
automated response.  The APT is complementary 
to these projects, however, as they are located in the Northern 
Hemisphere, and the APT is located in the Southern Hemisphere 
for access to the far southern sky. %

Conventional telescopes have been used to search for OTs,
however, the APT program has many advantages over these searches. 
Most telescopes suffer from scheduling problems and can therefore miss
bursts, they have a low event rate (mostly Sax alerts),
and they lack the installed and maintained software required to
identify OTs  on the same night as their discovery. 
 The latter capability is very important, in order that follow-up 
spectroscopy and other measurements may be made while the OT 
is near peak brightness.  The rapid, dedicated APT program avoids
these problems, and has an event rate more than six times that for
typical telescopes (due to its BATSE follow-up capability).  In addition,
the APT is immune to Sax malfunctions.  The most recent Sax problems
include gyro failure, which means delays in producing sub-arc minute 
imager positions.  Without the sub-arc minute positions, searches by
typical telescopes are much more time consuming.

\section*{Conclusion}

A significant-sized sample of OT light curves and more detailed 
follow-up observations (e.g. spectroscopy) would be likely 
to yield significant, rapid progress in this field.  The 
data would clarify controversies in the existing observations, 
such as host galaxy frequency
and possible variable extended emission.  A frequent 
lack of hosts would cause modifications to the 
neutron star - neutron star event scenario;  confirmation of 
variable extended emission would locate some GRBs near 
our own galaxy.  If the current trend continues, and $\sim$ 25\% 
of localized bursts yield OTs, our project's high event rate will
produce a sample of observations of 10 OTs after only 1-2 years of
operation.  

\subsection*{Figure Captions } 
Figure 1. Duration (T$_{90}$ = time to measure 90\% fluence) of GRBs in the
4B BATSE catalog; many are longer than 20 sec.  
Figure 2.  The 5$\sigma$ sensitivity 
of the APT with increasing delay (and integration) time after 
a GRB trigger.  The sensitivity of the APT is shown 
both truncated at 1 and 6 hours of co-added exposures; final 
verification of the sensitivity of our longest co-added exposures
is now in progress.  Figure 3. A 320 sec. image taken with
the APT at the position of GRB970326b.  The sensitivity is better than
V= 19.6 at 5 $\sigma$.  Many sources are present in the wide field.   %

\end{document}